\documentclass[%
twocolumn, superscriptaddress,
preprintnumbers,
amsmath, amssymb,
aps, pra
floatfix
]{revtex4-2}

\usepackage{graphicx}
\usepackage{dcolumn}
\usepackage{bm}

\usepackage{color}
\usepackage{hyperref}
\usepackage{comment}
\usepackage{makecell}

\usepackage{orcidlink}

\begin{document}

\title{Learning Kinetic Monte Carlo stochastic dynamics with Deep Generative Adversarial Networks}

\author{Daniele Lanzoni \orcidlink{0000-0002-1557-6411}}
\email{Contact author: daniele.lanzoni@unimib.it}
\affiliation{Department of Physics, University of Genova, Via Dodecaneso 33, 16146 Genova, Italy}
\affiliation{Department of Materials Science, University of Milano-Bicocca, Via R. Cozzi 55, I-20125 Milano, Italy}
\author{Olivier Pierre-Louis \orcidlink{0000-0003-4855-4822}}%
\affiliation{Institut Lumière Matière, UMR5306 Université Lyon 1—CNRS, 69622 Villeurbanne, France}%
\author{Roberto Bergamaschini \orcidlink{0000-0002-3686-2273}}
\affiliation{Department of Materials Science, University of Milano-Bicocca, Via R. Cozzi 55, I-20125 Milano, Italy}
\author{Francesco Montalenti \orcidlink{0000-0001-7854-8269}}
\affiliation{Department of Materials Science, University of Milano-Bicocca, Via R. Cozzi 55, I-20125 Milano, Italy}%

\date{\today}

\begin{abstract}
We show that Generative Adversarial Networks (GANs) may be fruitfully exploited to learn stochastic dynamics, surrogating traditional models while capturing thermal fluctuations. Specifically, we showcase the application to a two-dimensional, many-particle system, focusing on surface-step fluctuations and on the related time-dependent roughness. After the construction of a dataset based on Kinetic Monte Carlo simulations, a conditional GAN is trained to propagate stochastically the state of the system in time, allowing the generation of new sequences with a reduced computational cost. Modifications with respect to standard GANs, which facilitate convergence and increase accuracy, are discussed. The trained network is demonstrated to quantitatively reproduce equilibrium and kinetic properties, including scaling laws, with deviations of a few percent from the exact value. Extrapolation limits and future perspectives are critically discussed.
\end{abstract}

\maketitle

\section{Introduction}

Stochastic evolution laws are a key ingredient to simulate many processes of both fundamental and applied interest in condensed matter physics, materials science and engineering. Fluctuations are central for a proper description of non-equilibrium processes such as nucleation and growth, roughening of free surfaces, rare or activated events and phase transitions~\cite{tromp2002thermodynamics, battaile2002kinetic, debenedetti1996metastable, saito1996statistical, evans2006morphological}, to cite a few examples. Too coarse approximations of the involved probability distributions lead to inaccurate computational models that are unable to quantitatively reproduce experimental reality. On the other hand, more accurate approaches, such as molecular dynamics (MD)~\cite{hoyt2010fluctuations, shuichi1991constant} and Kinetic Monte Carlo (KMC)~\cite{voter2007introduction, henkelman2001long, xu2008adaptive}, may be limited in the possibility of reaching realistic temporal and spatial scales.

In recent years, machine learning (ML) has emerged as a new tool to infer probability distributions from data in several fields, allowing for the automatic extraction of correlations between observations~\cite{goodfellowdeep2016, mehta2019high}. Generative models in particular deal with the task of obtaining ML tools capable of generating new samples from a distribution given only samples extracted from it~\cite{harshvardhan2020comprehensive}. Among these approaches, Generative Adversarial Networks (GANs)~\cite{goodfellow2014generative} have already proven outstanding capabilities in several fields, e.g,. providing a versatile tool for the generation of images~\cite{karras2017progressive, han2018gan}, audio~\cite{kong2020hifi} and text~\cite{zhang2017adversarial}. Generated samples are often so similar to real ones to seem almost impossible to be distinguished. In recent years, they have also been successfully employed in several applications in condensed matter systems for accelerating ``static'' tasks. For instance, impressive results in the generation of microstructures~\cite{hsu2021microstructure, chun2020deep, nguyen2022synthesizing} and the extraction of configurations from equilibrium distributions~\cite{lee2025thermodynamic} have been reported. Considering also other fields of computational physics, promising applications may be found in the generation of protein folding~\cite{gupta2019feedback, repecka2021expanding} or astrophysical~\cite{schawinski2017generative} and particle physics~\cite{paganini2018accelerating} data, to cite a few. Unfortunately, in terms of time dependent problems, only low-dimensional cases has been shown to date. In particular, the non-linear dynamics of a single particle~\cite{yeo2022generative, stinis2024sdyn} and diffusion processes in 1D~\cite{lanzoni2023accurate} have been targeted. To our knowledge, no works have managed to tackle the stochastic evolution of large, many-particle systems. A first step in this direction was achieved for small molecules (around 20 atoms) using the ``score dynamics'' method in Ref.~\cite{hsu2024score}. There, Hsu et al. show that a diffusion model (a generative approach alternative to GANs) may be used to evolve in time the configuration of the system, effectively generating stochastic trajectories without the need, after training, of resorting again to molecular dynamics.

In the present work, we show that GANs may be applied to larger, many-particle condensed matter systems, focusing on the evolution of a crystal surface involving atomic step motion that is usually implemented by KMC simulations~\cite{evans2006morphological}. The advantages of the proposed method are threefold.

First, a speed-up in simulation times is obtained without a corresponding degradation of predictive capabilities. We show here an acceleration of $\approx 40\times$ with respect to KMC.

Second, symmetries and conservation laws can be easily considered by a suitable choice of the Neural Network architecture, leveraging strong physical priors to constrain the number of parameters and enhance the model accuracy. On the other hand, no detailed knowledge of the underlying potential energy landscape is required to train the model. In perspective, the approach may therefore be considered fully transferable to alternative stochastic evolution methods and, in principle, even experimental observations, if a suitable dataset is provided.

Third, the configurations of the system can be conveniently represented as images, leveraging the developments of deep learning in computer vision. This also allows a connection with other classes of methods for the mesoscale description of materials on a continuum level, such as phase field~\cite{steinbach2009phase, provatas2011phase}, a property that previous KMC acceleration schemes~\cite{shim2005semirigorous} do not share. Several attempts using ML approaches to replicate simulations of the temporal evolution of microstructures have been performed in recent years~\cite{Lanzoni2022PRM, wang2023machine, MontesdeOcaZapiain2021npj, oommen2024rethinking}, with computational speed-ups of orders of magnitude~\cite{fan2024accelerate, lanzoni2024extreme}. Most of the efforts, however, focus on approximating deterministic evolution laws, usually in the form of Partial Differential Equations (PDEs). Our work fills in the gap, allowing us to seamlessly insert thermal fluctuations and atomistic details in a continuum-like description.

The paper is organized as follows: first, we discuss the generalities of the KMC model used to generate the dataset and introduce the Generative Adversarial Network framework. Then, we specify the modifications the standard GAN training requires to obtain quantitative predictions from the generated dynamics. In Results and Discussion, we analyze the main outcomes of the study. In particular, both equilibrium and kinetic properties are considered, showing strong accordance with KMC statistical properties. Moreover, the generalization capabilities of the GAN method are tested, and examples of extrapolation behavior are shown. Finally, in the Conclusions, a discussion of possible future developments is provided.

\section{Methods} \label{sec::methods}

\subsection*{Kinetic Monte Carlo model and dataset construction}

Kinetic Monte Carlo is a method capable of simulating various physical processes and retaining an atomistic description of the system~\cite{bortz1975new, voter2007introduction}. KMC trajectories realize a Markov chain of states connected by stochastic transition rates. Given the state of the system at time $t$, the next configuration is sampled from the distribution of neighboring states such that the probability is proportional to the transition rate, which in turn is, in our case, related to an energy barrier height. Concurrently, following the Bortz-Kalos-Lebowitz (BKL) algorithm~\cite{bortz1975new}, the simulation time is advanced based on the total exit rate from the current state, which makes the method particularly effective in conditions dominated by activated processes. KMC can reach experimentally relevant spatial and temporal scales, and may efficiently simulate condensed matter systems of fundamental and applied interest~\cite{battaile2002kinetic, cheimarios2021monte, giesen2001step, evans2006morphological}.

As mentioned above, we here focus on the problem of the stochastic evolution of atomic steps on a crystalline surface. We have chosen this process for the present study for several reasons. First, since step morphology determines the density of preferential reaction and attachment sites, it is relevant in surface science for its impact on catalysis~\cite{yates1995surface, norskov2008nature} and crystal growth~\cite{giesen2001step, cheimarios2021monte}. Among the different evolution mechanisms, the specific case of edge diffusion is considered, which is known to lead to challengingly slow evolutions and to be an accurate model for the dynamics of atomic steps at the surface of metals~\cite{giesen2001step, evans2006morphological} below the surface roughening transition~\cite{saito1996statistical}. Second, it allows for showcasing how GANs may be employed to learn a stochastic evolution law in conditions where the dynamics is dominated by fluctuations: step roughness is present at any finite temperature~\cite{saito1996statistical} while net motion is absent. Finally, this case is also particularly interesting as it required the addition of the mass conservation laws into the GAN approach.

\begin{figure}
    \centering
    \includegraphics[width=\linewidth]{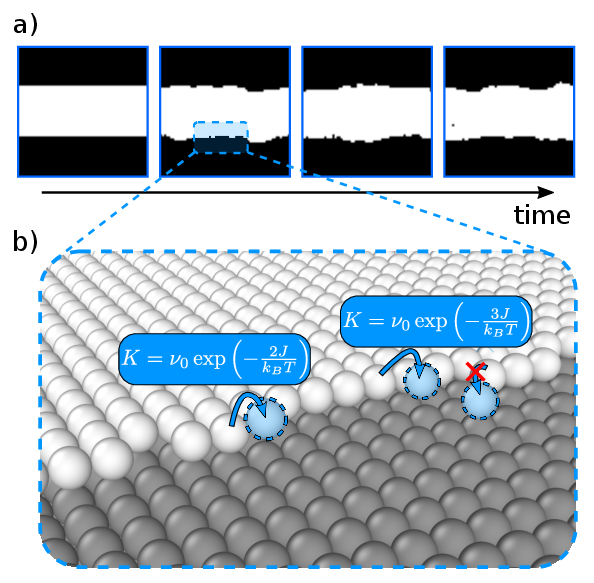}
    \caption{(a) Sample of an evolution obtained by KMC simulation. (b) Schematics of the KMC rules for jumping rate calculations. Some of the possible movements and corresponding rates are reported. The crossed-out move is not possible, as it would lead to an adatom detachment and the violation of edge diffusion conditions.}
    \label{fig::KMC_scheme}
\end{figure}

To generate a dataset of random trajectories, we use 2D lattice KMC simulations of a $\{ 100 \}$ surface of a simple cubic material with lattice parameter $a=1$.  The initial state is set to a stripe-shaped domain, which progressively roughens due to thermal fluctuations, as shown in Fig.~\ref{fig::KMC_scheme}(a). Black and white pixels represent occupied and unoccupied sites, respectively. Periodic boundary conditions (PBCs) are used. One main advantage of this geometry is the availability of analytical predictions for the stochastic evolution of the steps and their roughness~\cite{misbah2010crystal, khare1998unified}, as also discussed in the next section, that can be checked quantitatively.

Atoms move by edge diffusion~\cite{lai2019reshaping, giesen2001step}, conserving their total number: a particle may jump to a neighboring site (first and second neighbors are considered) only if it is unoccupied but also has at least one occupied first neighbor site. This rule ensures that diffusion moves only happen at step edges (hence "edge diffusion") and that no atom can detach and become an independent adatom on the substrate. For the present work, the jump rate $K$ is calculated based on a simple first-neighbor-only interaction approximation:
\begin{equation}
    K = \nu_0 \exp{\left( -\frac{nJ}{k_BT} \right)} ,
\end{equation}
with $\nu_0$ being the jumping attempt frequency, $n$ the number of first neighbors, $J$ the bond energy parameter, $k_B$ is the Boltzmann constant and $T$ the absolute temperature. A sketch of the underlying physical model is reported in Fig.~\ref{fig::KMC_scheme}(b). This model has been chosen over more refined schemes, such as those in Refs.~\cite{henkelman2001long, xu2008adaptive}, because it is the simplest one that captures the non-trivial character of edge dynamics.

The training set is composed of $300$ independent KMC runs with $J=0.1$~eV and $T=290$~K, so that $k_BT/J \approx 0.25$. The computational domain is a $64 \times 64$ lattice, with the stripe having an initial uniform thickness of $25$ atoms. For each simulation, a total of $1000$ subsequent snapshots are saved at regular time intervals $\tau$, whose role will become clear in the next section. In units of $1/\nu_0$, $\tau = 5\times10^4$. This way, subsequent states are separated by several thousand diffusion steps. The choice of $\tau$ is somewhat arbitrary, but it needs to meet some straightforward requisites. For instance, if a too short time interval is chosen, the system is not evolving significantly within $\tau$ and the acceleration provided by the GAN approach will be small. At the other extreme, if $\tau$ is so long that it is comparable to the system equilibration time, the task of learning the transient dynamics from the stripe to a rough state would be impossible and the GAN would simply sample from the equilibrium distribution.

\begin{figure*}
    \centering
    \includegraphics[width=\textwidth]{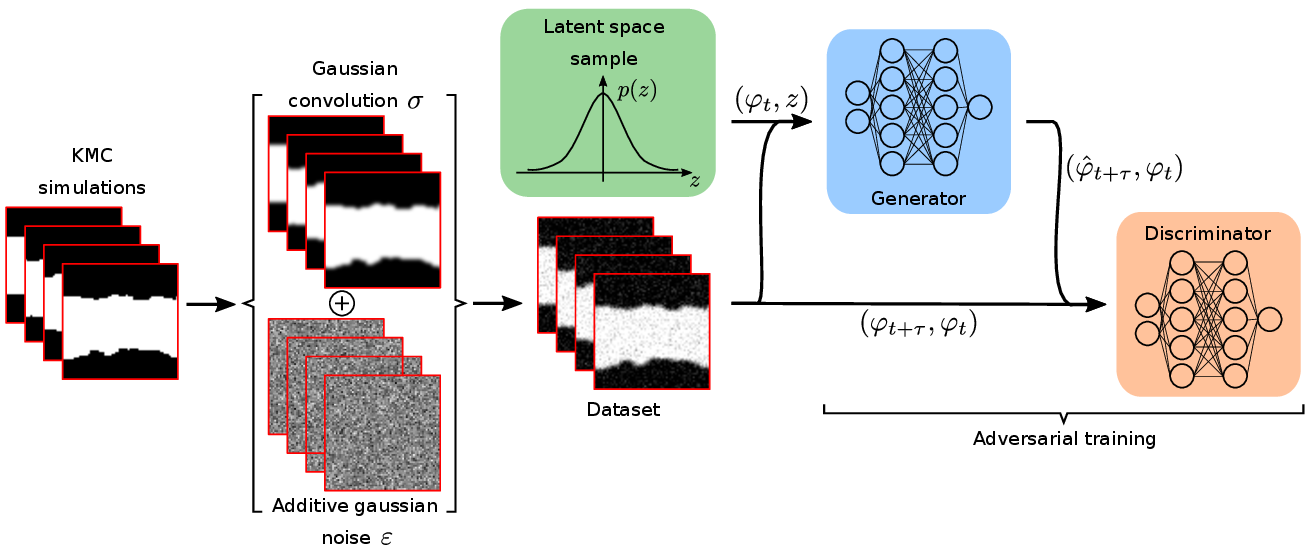}
    \caption{GAN training workflow used in the present work. The dataset is built upon KMC simulations, to which gaussian convolution and additive random noise $\varepsilon$ are applied. It is then used for the adversarial training procedure: the Generator takes the current state of the system $\varphi_t$ and a latent space sample $z$ as inputs and outputs the subsequent state $\hat\varphi_{t+\tau}$. The Discriminator tries to identify which of the couples $(\varphi_{t+\tau}, \varphi_t)$ and $(\hat\varphi_{t+\tau}, \varphi_t)$ comes from the dataset and which comes from the Generator.}
    \label{fig::GAN_scheme}
\end{figure*}

\subsection*{Analytical predictions}

As long as the substrate is below its surface roughening transition temperature, an initially flat stripe progressively increases its corrugation in time, up to reaching a stationary value~\cite{saito1996statistical}. Since direct comparison of stochastic trajectories is not meaningful, it is necessary to resort to average quantities, expectation values and correlations to assess the quality of the trained NN models. In the following, atomic step roughness will be the main quantity used to perform this analysis.

Analytical expressions for the roughening dynamics are available in the continuum small-slope limit. Given a configuration with step boundary traced by $h(x)$ ($x \in [0, L]$, assuming PBCs), $W^2$ is defined as the variance
\begin{equation} \label{eq::roughness_definition}
    W^2 = \frac{1}{L} \int_0^L (h(x) - \bar{h})^2 dx,
\end{equation}
where $\bar{h}$ is the average step-edge position $\int_0^L h(x) dx/L$.

For individual trajectories, $W^2$ fluctuates randomly, but time evolution of the \textit{ensemble} average of the roughness, here indicated with $\langle W^2 \rangle$, has an analytical expression provided by the infinite series~\cite{marguet2022interface}:
\begin{equation}  \label{eq::time_roughness_noconv}
    \langle W^2(t) \rangle = \frac{L k_B T}{12 \tilde{\gamma}} \frac{6}{\pi^2} \sum_{n>0} \frac{1-\exp{\left[ -2(2 \pi n)^{4} u \right]}}{n^2},
\end{equation}
with $u= {\tilde{\gamma} M t}/{k_B T L^{4}}$ and $M$ is a mobility constant that may be related to the KMC jumping rates. Notice that $\langle W^2 \rangle$ contains the same information as the usual time-correlation function~\cite{misbah2010crystal}. In the low temperature regime~\cite{pierre2001continuum, bartelt1992equilibration}, the following approximation holds
\begin{equation} \label{eq::mobility}
    M \approx M_\text{th} = \nu_0 a^5 \exp{\left( \frac{-2J}{k_B T} \right)}.
\end{equation}
Using the parameters reported in the previous section for KMC simulations, we obtain that $M_\text{th} = 3.34 \times 10^{-4}$.

The quantity $\tilde{\gamma}$ is instead the so-called ``edge stiffness'', which in the used model may be expressed as~\cite{marguet2022interface}:
\begin{equation} \label{eq::stiffness}
    \tilde{\gamma} = \frac{k_BT}{2 a} \left( \varepsilon^{-1/4} - \varepsilon^{1/4} \right)^2,
\end{equation}
where $\varepsilon = \exp{ \left( - J/k_BT \right)}$.

Irrespective of the value of $M$, the ensemble average of roughness reaches an equilibrium value $\langle W^2 \rangle_\text{eq}$ for $t \to \infty$~\cite{saito1996statistical},:
\begin{equation} \label{eq::eq_roughness_noconv}
    \langle W^2 \rangle_\text{eq} = \frac{k_B T L}{12 \tilde{\gamma}}.
\end{equation}

\subsection*{Stochastic dynamics via Generative Adversarial Networks}

Given the KMC model outlined in the previous section, the states of the system may be conveniently represented as a sequence of binary images. This is particularly useful, as it allows one to import methodologies based on convolutional NN (CNN)~\cite{bishop2006pattern, goodfellowdeep2016} and other computer vision concepts in the Generative Adversarial Network approach used in this work. Importantly, this is not only an advantage in terms of implementation and computational costs, but also automatically handles the translational symmetry of the system, since CNNs~\cite{goodfellowdeep2016} are equivariant with respect to image translations. Moreover, the binary image representation is also explicitly symmetric to particle exchange, thus removing a second, known symmetry that the NN would need to learn. Another critical advantage of NN composed only of convolutions and pointwise non-linearities (so-called ``fully-convolutional'' networks) stems from the possibility of operating on inputs of arbitrary sizes. This means that, after training, the method will be able to evolve states on larger (and smaller) domains, allowing for spatial-size generalization.

In general, GANs training may be framed as an adversarial game~\cite{goodfellow2014generative} between two NN, called the Generator $G$ and the Discriminator $D$. $G$ has to map samples $z$ (called latent vectors) from a known, well-behaved distribution to samples that resemble those in the dataset. At the same time, $D$ tries to classify both the generated samples $G(z)$ and those from the dataset, with the aim of distinguishing real and generated data. As in other Deep Learning approaches, training is performed through the minimization of loss functions $\mathcal{L}$. In recent years, several variants of GANs have been proposed with different objectives~\cite{mao2017least, arjovsky2017wasserstein}. In this work, however, we use the original formulation from Goodfellow et al.~\cite{goodfellow2014generative}, since from several tests it yields satisfactory results. The Generator and Discriminator loss functions, respectively, read:
\begin{equation}
\begin{split}
    & \mathcal{L}_G = -\mathbb{E}[\log{\hat{s}}];\\
    & \mathcal{L}_D = -\frac{1}{2}(\mathbb{E}[\log{s}] + \mathbb{E}[\log{(1-\hat{s})}]),
\end{split}
\end{equation}
where $\mathbb{E}$ is the expectation value operator. The values $s$ and $\hat{s}$ are the ``scores'' between $0$ and $1$ that $D$ assigns to real and generated samples. A higher score corresponds to a higher confidence that the sample comes from the dataset and not from $G$. Indicating an element of the training set with $\varphi$, and the corresponding Generator sample as $\hat{\varphi}$, then
\begin{equation}
\begin{split}
    & s = D(\varphi_t);\\
    &\hat{s} = D(\hat{\varphi}_t) = D(G(z)).
\end{split}
\end{equation}
Convergence is obtained when $s=\hat{s}=1/2$ and both loss functions are equal to $\log 2$ (Nash equilibrium).

We exploit the GAN formalism to effectively learn stochastic dynamics in the following way. Given the state of the system $\varphi_t$ at time $t$, the core idea is to use the Generator to implicitly learn the conditional probability distribution of states after a fixed time interval $\tau$, $\mathbb{P}(\varphi_{t+\tau}|\varphi_t)$. This can be done by augmenting the input of $G$ to also contain $\varphi_t$, i.e., using the framework of conditional GANs~\cite{mirza2014conditional}. In practice, since the Generator is implemented as a CNN and $z$ is chosen to be an $N$-channel image ($N=30$ yielded the best results in our case), the input of $G$ is composed through concatenation of $z$ and $\varphi_t$, i.e. a $(N+1)$-channel image. A new sample for the next state of the system is therefore obtained as:
\begin{equation}
    \hat{\varphi}_{t+\tau} = G(\varphi_t, z).
\end{equation}
Similarly, $D$ scores are also conditioned on the previous state of the system, and its input is also modified so that it is constituted by the couples $(\varphi_{t+\tau}, \varphi_t)$ and $(\hat{\varphi}_{t+\tau}, \varphi_t)$. After training, an individual stochastic trajectory of the system is obtained as a Markov chain by operating the Generator as an auto-regressive model as in our previous work at Ref.~\cite{lanzoni2023accurate}, i.e., by re-feeding its own output as a new input. Notice that, at variance with the BKL algorithm, the states in the chain are not separated by a single elementary transition, but by the fixed time interval $\tau$, during which several thousand diffusion events have happened.

We now report the general $G$ architecture used in this study for the interested reader. A 5-layer Convolutional NN (CNN) is employed. In particular, a ResNet-like architecture~\cite{he2016deep} was used, with 10 channels for each hidden layer, amounting to $\approx 30000$ training parameters. To better capture time correlations, the Generator is set to learn the (stochastic) residual $\delta \hat{\varphi} = \hat \varphi_{t+\tau} - \varphi_t$ instead of generating the next state directly, a strategy that has already proven effective in deterministic contexts~\cite{lanzoni2024extreme, fan2024accelerate} and has shown to yield better results also in our tests. As already discussed, this architecture is also particularly convenient to ensure exact conservation of the number of particles: to this aim, it is sufficient to remove the spatial average value of $\delta \hat{\varphi}$ before calculating the output $\hat{\varphi}_{t+\tau} = \varphi_t + \delta \hat{\varphi}$,.

Another hyperparameter is the dimensionality of the latent vector $z$. In principle, its actual value is not important, as long as it is larger than the dimensionality of the data manifold~\cite{arjovsky2017towards}. On the other hand, the larger the number of $z$ components, the higher the computational cost. Leaving the comprehensive study of this hyperparameter for future work, we found after some tests that a $z$ extracted from a multivariate gaussian with the same spatial dimensionality as $\varphi_t$ and $N = 30$ channels (i.e. $30 \times 64 \times 64$ floating point numbers in training) yields satisfactory results without increasing computational costs too much.

For the Discriminator, a similar convolutional ResNet architecture is used. Since the output of $D$ is required to be a scalar, however, these are interleaved with max pooling layers~\cite{goodfellowdeep2016} with a $2 \times 2$ receptive field each time halving the feature map, for a total of $\approx 28000$ parameters.

\subsection*{Adapting GANs to KMC}

One of the main issues of GANs is their notorious instability during training~\cite{mescheder2018training}. Indeed, direct application of the approach described in the previous section would yield poor results in terms of convergence to the Nash equilibrium and generation of trajectories. One of the reasons has already been studied in depth in Ref.~\cite{arjovsky2017towards}. There, Arjovsky and Bottou have proven that training is unstable if the intersection between the support of the data and the generated probability distribution has zero measure. This condition happens if the data are not absolutely continuous variables, as in the case of the \textit{discrete} KMC configurations. By construction, the set of binary images used here has zero measure in the set of $G$ output, which are real-valued $64 \times 64$ arrays. GAN training is therefore expected to be unstable.

The problem can be solved with two modifications. First, as already proposed in Ref.~\cite{arjovsky2017towards}, adding white gaussian noise $\varepsilon$ to both generated and true data stabilizes training by expanding the support of probability distributions and removing possible degeneracies. A similar technique is used in the regularization approach called noise injection~\cite{bishop1995training}. This strategy has already been shown to be effective in the case of training on 1D, discrete-valued random walks from KMC simulations~\cite{lanzoni2023accurate}. In the same work, we found that the standard deviation of the additive noise should be $\approx 0.25 \times$ the Euclidean distance between the underlying discrete values in the data. Since $\varphi$ admits only two values in the binary-image representation used in this work, a standard deviation of $0.25$ should therefore be used. Unfortunately, this level of noise may hide important features in the dynamics. A smaller-variance additive noise could however be used if intermediate values between $0$ and $1$ were present in $\varphi$.

This is attained here by convolving the state of the system with a gaussian function. A discrete approximation based on a $5 \times 5$ kernel for the gaussian convolution has been used to match the pixelization, with standard deviation $\sigma \approx 1.11$. Notice that this approach does not eliminate the need for stochastic random noise injection, as the convolution is computed on binary images and a fixed grid, still providing states with discrete values. Nonetheless, the spacing in $\varphi$ values is now smaller than $1$. More specifically, considering the initial stripe configuration, after the convolution the maximal separation in $\varphi$ for adjacent pixels is $\approx0.376$. The $0.25\times$ rule of thumb hence allows for random noise with a standard deviation as small as $\text{std}(\varepsilon) = 0.1$, which indeed yields stable training, as shown in the next Section. A schematic representation of the training procedure, also involving the specialized steps required to obtain convergence, is reported in Fig.~\ref{fig::GAN_scheme}.

The convolution procedure, akin to a gaussian local average, has the second important effect of smearing the ``sharp-interface'' details of $\varphi$ into a ``diffuse-interface'' representation. This allows for a natural connection between the original atomistic description of the system with continuum models, such as phase-field approaches~\cite{steinbach2009phase, provatas2011phase}. These represent an important class of simulation tools for surface and nanoscale physics~\cite{bergamaschini2016continuum, li2009geometric, provatas2011phase}, which surpass some of the limitations of atomistic models in terms of accessible time and spatial scales. On the other hand, the addition of some details, such as anisotropies and thermal fluctuations, which are inherently contained in the GAN approach presented in this work, often requires significant computational and theoretical efforts.

\begin{figure*}
    \centering
    \includegraphics[width=\textwidth]{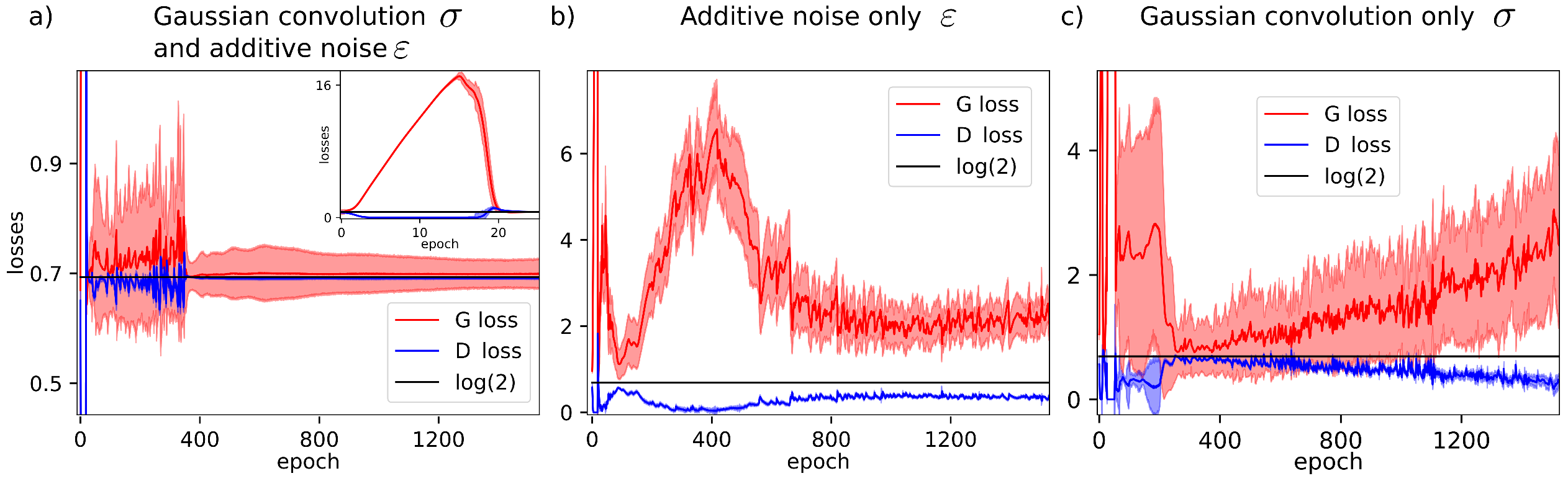}
    \caption{Comparison between losses for models trained using the specialized noise injection and gaussian convolution procedures in different combinations. Notice the difference in scale for the $y$-axis for different panels. Solid lines correspond to epoch averages of $\mathcal{L}_G$ (red) and $\mathcal{L}_D$ (blue), while shaded areas are $2$ standard deviations in fluctuations calculated on each epoch. The Nash $\log \left( 2 \right)$ value is reported as a black line. (a) Training using both gaussian convolution and additive noise; (b) training using additive noise only; (c) training using gaussian convolution only. Inset in panel (a) shows the initial spike in the loss functions (epochs $\approx 1-20$).}
    \label{fig::loss_function}
\end{figure*}

Going from a sharp to a diffuse-interface representation, however, has its costs: if convoluted $\varphi$ are used to evaluate the step roughness, short-wavelength details are lost, affecting both the equilibrium and time-dependent ensemble averages. Luckily, analytical expressions are available for both situations. The 2D convoution on $\varphi$ can be shown to be well approximated by a 1D convolution on $h(x)$ (see Appendix~\ref{sec::convolution_appendix}), implicitly defined by the $\varphi=1/2$ isoline. The time evolution law (Eq.~\ref{eq::time_roughness_noconv}) for the roughness of the convolved profile becomes:
\begin{equation}  \label{eq::time_roughness}
    \langle W_\sigma^2(t) \rangle = \frac{L k_B T}{12 \tilde{\gamma}} \frac{6}{\pi^2} \sum_{n>0} \frac{e^{-(2 \pi n)^2 \frac{\sigma^2}{L^2}}}{n^2} \left( 1-e^{-2(2 \pi n)^4 u} \right),
\end{equation}
where the $\sigma$ is the standard deviation of the gaussian kernel. All other symbols have the same meaning as in previous equations. At equilibrium ($\lim t \to \infty$), Eq.~\ref{eq::time_roughness} simplifies to~\cite{gagliardi2022controlling}:
\begin{equation} \label{eq::eq_roughness}
    \langle W^2_\sigma \rangle_\text{eq} = \langle W^2 \rangle_\text{eq} S(\sigma/L),
\end{equation}
i.e. the equilibrium value for $\langle W^2_\sigma \rangle$ is modified only by the multiplicative factor $S(\sigma/L)$
\begin{equation}
    S(\sigma/L) = \frac{6}{\pi^2} \sum_{n>0} \frac{1}{n^2} \exp{-\frac{n^2\sigma^24\pi^2}{L^2}},
\end{equation}
which is a function of the ratio between the gaussian kernel standard deviation and the domain size $L$. For $L=64$ and $\sigma \approx 1.11$, the correction factor for the training set is $S(\sigma/L) \approx 0.88$.

A last major issue in using GANs to generate stochastic trajectories stems from the fact that oscillations with respect to Nash equilibrium are usually persistent due to the stochastic and adversarial nature of the training process. Moreover, even if the Generator with the training loss $\mathcal{L}_G$ closest to the ideal Nash value is selected, small systematic biases could be present. When the model is operated autoregressively, these may accumulate and severely affect the generated dynamics. These problems have already been addressed and may be effectively mitigated using the multi-model approach proposed in Ref.~\cite{lanzoni2023accurate}: instead of using a single $G$, it is possible to construct an ensemble of models and generate samples from the corresponding average distribution. The ensemble is constructed using the last $M$ Generators obtained in a single training, i.e. those having a loss function oscillating around the Nash value. During trajectory generation, at every timestep one of the $G$s is chosen with uniform probability from the ensemble, thus reducing the accumulation of systematic approximation errors~\cite{lanzoni2023accurate}. Notice that no training computational overhead is introduced by this procedure, as the ensemble is automatically formed during the usual GAN training procedure. For the present work, $M=1000$ has been chosen.

\section{Results and discussion} \label{sec::results}

\subsection*{Training and acceleration capabilities}

The GAN training workflow outlined in the previous section has been implemented in PyTorch~\cite{paszkepytorch2019}. Since identifying the hyperparameters in GAN training may be particularly challenging, we list here the main ones. Further details on specific implementations may be found in the provided source code. The batch size used was $1024$ and optimization was performed using RMSProp~\cite{hinton2012neural}, with parameters $\beta_1=0.1$, $\beta_2=0.99$ and a learning rate of $5 \times 10^{-6}$ for a total of $2000$ epochs. No weight decay or norm regularization on NN parameters was employed. Data augmentation was used to enforce symmetries not already present in the NN architecture, i.e., 90° rotations and mirror reflections have been applied.

After the typical initial spike in $\mathcal{L}_G$ (see inset of Fig.~\ref{fig::loss_function}(a)), the models rapidly approach the Nash condition, with persistent small fluctuations around the equilibrium as expected and shown in Fig.~\ref{fig::loss_function}(a). To explicitly show that the joint use of noise injection and convolution is necessary for training convergence, we also report a comparison with trainings performed where one of the procedures is removed in panels (b) and (c). Despite temporarily approaching the ideal $\log 2$ value during the initial stages of training, losses eventually drift away from the Nash equilibrium in both situations.

In terms of simulation acceleration capabilities, the generation via GAN of an individual full evolution of $1000$ steps ($\tau$ units of time apart each) for a $64 \times 64$ domain requires approximately $5$~s on GPU (Nvidia RTX A5000) and $\approx 9$~s on CPU, compared with the $\approx 3$~min necessary for the efficient in-house FORTRAN code used to generate the edge-diffusion dataset on the same CPU, yielding a computational speed-up of almost $40 \times$. While much higher acceleration factors are present in the literature for deterministic dynamics~\cite{fan2024accelerate}, it should be remembered that learning stochastic evolution is also a much more challenging task. Additionally, the model considered here implemented a computationally cheap bond counting rule for jumping rates. Advanced KMC models such as adaptive KMC~\cite{xu2008adaptive} and more realistic system-specific models~\cite{lai2019reshaping}, with more complex activated processes, have a higher computational cost. In contrast, we expect these additional complexities to have no or marginal impacts on GAN generation time.

\subsection*{Roughness evolution and fluctuations}

\begin{figure}
    \centering
    \includegraphics[width=\columnwidth]{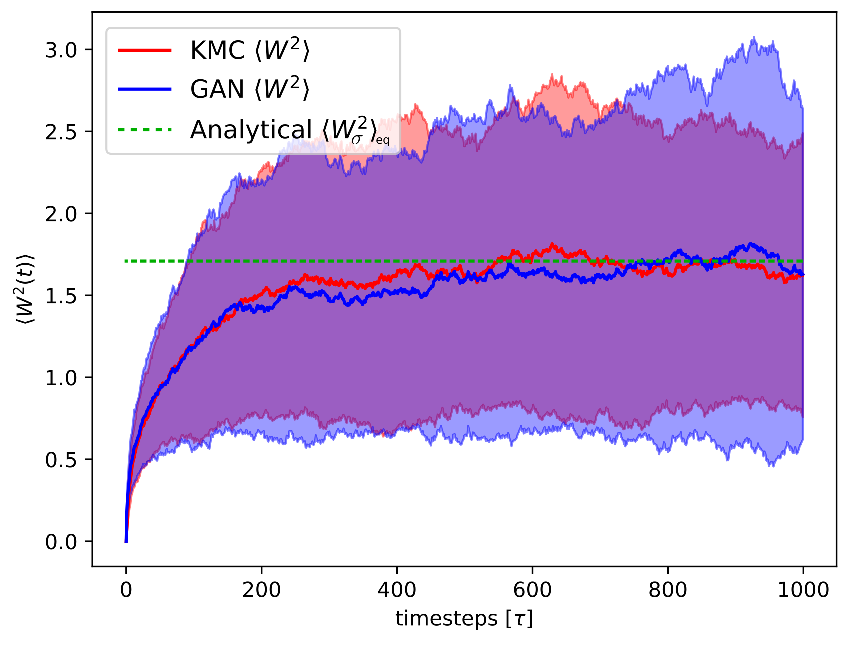}
    \caption{Comparison between the roughness evolution from KMC (red solid line) and GAN (blue solid line), as obtained from 300 independent edge-diffusion simulations. The analytical expected value at equilibrium $\langle W_\sigma^2\rangle_\text{eq}$ is reported as a dashed green line. Shaded areas are bounded by $\pm$ one standard deviation, evaluated on the 300 independent runs.}
    \label{fig::roughness_evo}
\end{figure}

A comparison between the roughness coming from KMC and that of GAN simulations, together with the theoretical saturation value of Eq.~\ref{eq::eq_roughness}, accounting for the convolution correction factor, is reported in Fig.~\ref{fig::roughness_evo} (red, blue, and green lines respectively). In order to perform a fair comparison with the KMC dataset, a total of 300 trajectories were generated using the trained Generator. The shaded area corresponds to one standard deviation in the roughness value, representing ensemble fluctuations. It may be observed that the GAN approach not only catches the overall qualitative time-dependent behavior but also achieves quantitative accuracy.

To be more precise, we first compare equilibrium properties from KMC and GAN evolutions. The values of $\langle W^2 \rangle_\text{eq}$ are $1.69 \pm 0.05$ and $1.67 \pm 0.06$ respectively, as also reported in Table~\ref{tab::roughness_eq}. The confidence interval corresponds to one standard deviation. This is calculated considering the variance of $\langle W^2 \rangle$ in the last half of the reported evolution, where the equilibrium state has already been reached. Accordance is remarkable, given the strength of typical fluctuations of this quantity in the simulations, with an absolute deviation between KMC and GAN predictions of $\approx 0.8\%$. Both values fall within one standard deviation from the analytical prediction of $1.71$.

As an alternative estimation of the prediction error, it is interesting to discuss the equilibrium roughness deviation observed in terms of physical parameters. Based on Eq.~\ref{eq::eq_roughness}, should the difference with the theoretical value be attributed to an error in the inference of the bonding parameter $J$, then this would correspond to $J_\text{GAN}=0.096$~eV (vs the expected $0.1$~eV). If instead the deviation is attributed to temperature, $T_\text{GAN} = 301.3$~K, to be compared with the $300$~K used as an input to generate the dataset. The relative errors are both $\approx 0.4\%$, which proves that the GAN can be reliably used to predict physical parameters with high accuracy.

\begin{table}
    \centering
    \begin{tabular}{ | c | c | c | c |}
        \hline
        Quantity & KMC & GAN & Analytical \\
        \hline
        $\langle W^2_\sigma \rangle_\text{eq}$ & \makecell{$1.69$ \\ $\pm 0.05$} & \makecell{$1.67$ \\ $\pm 0.06$} & $1.71$ \\
        \hline
        $\text{std}(W^2)_\text{eq}/\langle W^2_\sigma \rangle_\text{eq}$ & \makecell{$0.53$ \\ $\pm 0.03$} & \makecell{$0.63$ \\ $\pm0.04$} & $\sqrt{2/5} \approx 0.6324$ \\
        \hline
        $M$ & $3.27 \times 10^{-4}$ & $3.00 \times 10^{-4}$ & $3.34 \times 10^{-4}$ \\
        \hline
        $\alpha$ & $-$ & $3.08\times10^{-2}$ & $\approx 3.01\times 10^{-2}$ \\
        \hline
    \end{tabular}
    \caption{Comparison between different properties as estimated from KMC and GAN simulations, together with their analytical reference value: $\langle W^2_\sigma \rangle_\text{eq}$ is the equilibrium value of roughness (accounting for convolution effects), $\text{std}(W^2)_\text{eq}/\langle W^2_\sigma \rangle_\text{eq}$ is the ratio between equilibrium roughness standard deviation and its mean value, $M$ is the step edge mobility constant and $\alpha$ the proportionality factor between equilibrium roughness and step length. Where reported, confidence intervals are indicated by $\pm$.}
    \label{tab::roughness_eq}
\end{table}

Higher moments of the fluctuations, such as the variance of the roughness, are more challenging to measure. In the continuum limit, the ratio between the standard deviation of $\langle W^2 \rangle_\text{eq}$ and its mean value is expected to be $\sqrt{2/5} \approx 0.6324$ (see Appendix~\ref{sec::roughness_derivation}). The estimated values from simulations are $0.53\pm0.03$ (KMC) and $0.63\pm0.04$ (GAN). This not only confirms the capability of the GAN approach in reproducing fluctuations, but is also a confirmation that the model is not exhibiting one of the typical problems in Generative Adversarial Networks, i.e., mode collapse~\cite{arjovsky2017towards}. This is a phenomenon in which the Generator provides good samples, but the distribution is almost degenerate and the variability of outputs is not representative of that present in real data. In contrast, the generated trajectories here fully recover the correct amplitude of the fluctuations.

While the possibility of obtaining equilibrium values is important, a key advantage in using the conditional GAN scheme is the possibility of recovering kinetic properties. In this respect, it may be observed that the correct qualitative behavior has indeed been captured. In order to quantify the prediction accuracy of kinetic properties, it is possible to extract the corresponding mobility constant from the initial transient of the dynamics by fitting Eq.~\ref{eq::time_roughness} with non-linear regression. The value obtained from KMC is $M_\text{KMC} = 3.27 \times 10^{-4}$, which is in substantial agreement with the theoretically expected value $M_\text{th} = 3.34 \times 10^{-4}$ of Eq.~\ref{eq::mobility}. From GAN simulations, $M_\text{GAN} = 3.00 \times 10^{-4}$, which has a relative error of $\approx 8.1\%$ with respect to the KMC value.

\subsection*{Relaxation dynamics from undulated step configurations}

\begin{figure}
    \centering
    \includegraphics[width=\columnwidth]{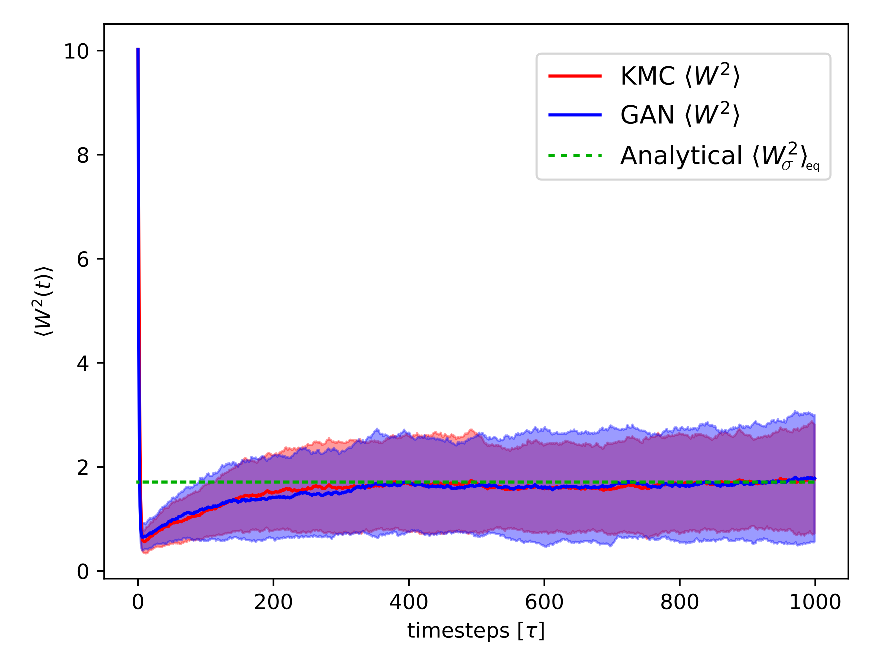}
    \caption{Comparison between GAN and KMC predictions (blue and red solid lines respectively) during relaxation dynamics from an initially undulated stripe configuration (amplitude $5$ lattice units). The dashed green line reports the analytical equilibrium roughness value. Domain length $L=64$ and perturbation wavevector $k=8\pi/L$.}
    \label{fig::relaxation64}
\end{figure}

Now that the Generator capabilities in reproducing statistical properties in conditions similar to those encountered in training has been assessed, we start exploring its generalization capabilities. As a first test, we consider the evolution of $\langle W^2 \rangle$ with an initial condition outside the distribution of the training dataset. Specifically, the stripe boundaries are now described by $A\sin{[kx + \phi]}$, with $A=5$ lattice units, $L=64$, $k=8\pi/L$ (hence $4$ periods in the computational cell), and phase factors $\phi$ chosen independently at random in $[0,2\pi]$. Fig.~\ref{fig::relaxation64} reports the roughness behavior as predicted from KMC and GAN, using an ensemble of $300$ independent trajectories for both, as in the previous analysis. Close accordance between the two curves may be observed: both approaches predict a fast initial drop of roughness, due to the decay of the amplitude of the initial single mode, leading to an ``overshooting'' of the equilibrium value, followed by a slower increase up to $\langle W^2_\sigma \rangle_\text{eq}$. This supports the hypothesis that the GAN predictive capabilities do not depend on having an initial condition identical to that of the training set.

\subsection*{Scaling properties}

\begin{figure}
    \centering
    \includegraphics[width=\columnwidth]{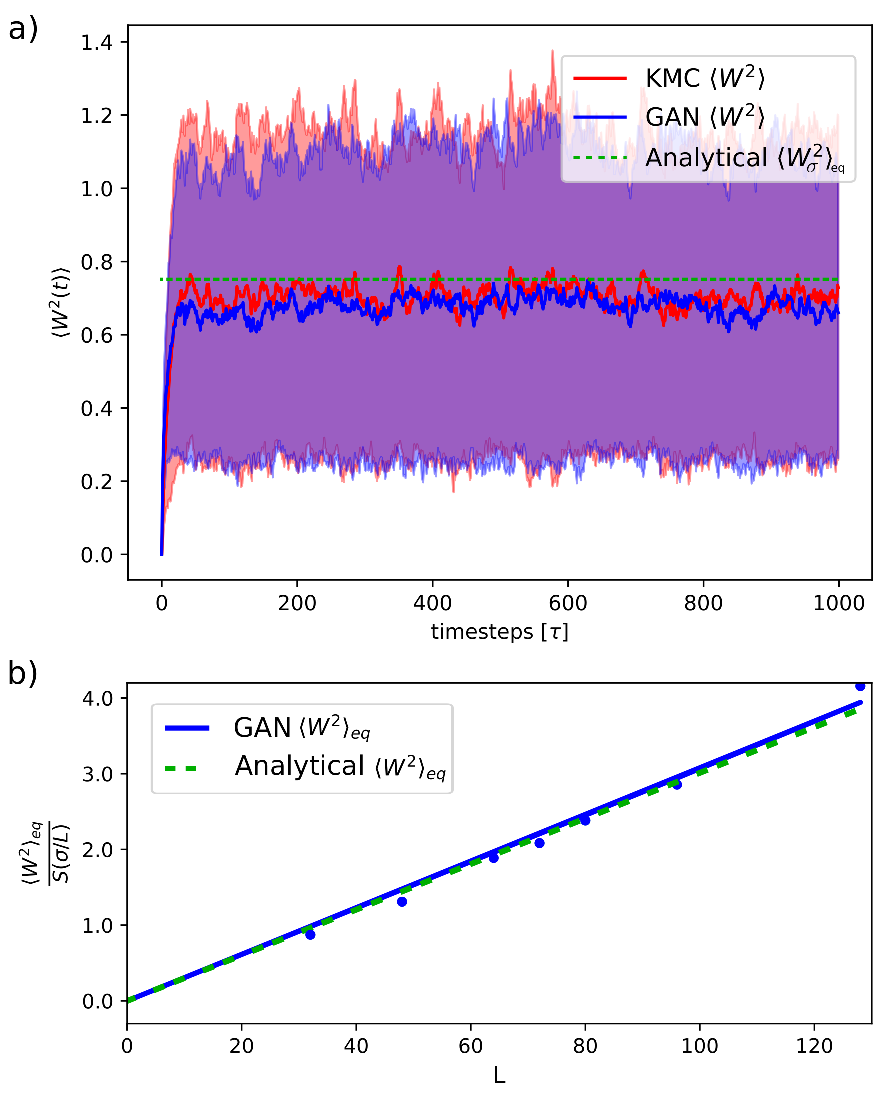}
    \caption{(a) Roughness evolution on a shorter computational cell ($32\times64$). Solid lines represent ensemble averages over 300 independent simulations (red for KMC and blue for GAN, respectively). The boundaries of the shaded areas represent $\pm$ one standard deviation intervals. (b) Scaling behavior of roughness as a function of domain length $L$ as predicted from theory (dashed green curve) and GAN simulations (blue curve). Error bars representing confidence intervals are not reported since they are smaller than the point size. The NN model is correctly capturing the linear scaling behavior.}
    \label{fig::scaling}
\end{figure}

We now test $G$ generalization capabilities further with respect to domain size. This is possible thanks to the fully-convolutional architecture used in $G$, as discussed in the Methods section. As a simple comparison, we show in Fig.~\ref{fig::scaling}(a) the time-dependent roughness as predicted from KMC and GAN for a smaller domain with respect to the training set one. Specifically, a $32\times64$ computational cell was used, thus halving the length of the stripe. Quantitative agreement can once again be observed, both in terms of ensemble averages and fluctuations.

As a more quantitative proof of the NN generalization capabilities, which also involves domains larger than those used in training, we perform an analysis of the scaling behavior of the equilibrium roughness as a function of the stripe size. From Eq.~\ref{eq::eq_roughness}, the theoretical value of $\langle W_\sigma^2 \rangle_\text{eq}/S(\sigma/L)$ should be linearly dependent on the stripe length $L$, with a proportionality constant $\alpha$ (using lattice units):
\begin{equation} \label{eq::scaling}
    \alpha = \frac{1}{6 \left( \varepsilon^{-1/4} - \varepsilon^{1/4} \right)^2}.
\end{equation}
With physical parameters reported previously, $\alpha \approx 3.01\times 10^{-2}$.

This quantity may be easily extracted for GAN predictions by least squares regression, as shown in Fig.~\ref{fig::scaling}(b). Each one of the points at different $L$ values is obtained by averaging the roughness value of $300$ independent trajectories once they reach equilibrium, as already discussed for the $64 \times 64$ case. For larger values of $L$, the longer time required to reach $\langle W^2 \rangle$ saturation has been accounted for by performing longer simulations ($1500 \tau$ for $L=96$ and $3000 \tau$ for $L=128$). This leads to slight differences in accuracy, but in all cases confidence intervals on roughness are below $4.5\%$ of their values. The initial condition used in this case is always a flat stripe. The estimated value from GAN simulation, considering zero intercept constraint, is $\alpha_\text{GAN} = 3.08 \times 10^{-2}$, with a relative error with respect to Eq.~\ref{eq::scaling} of $\approx 2.2\%$.

\begin{figure}
    \centering
    \includegraphics[width=\columnwidth]{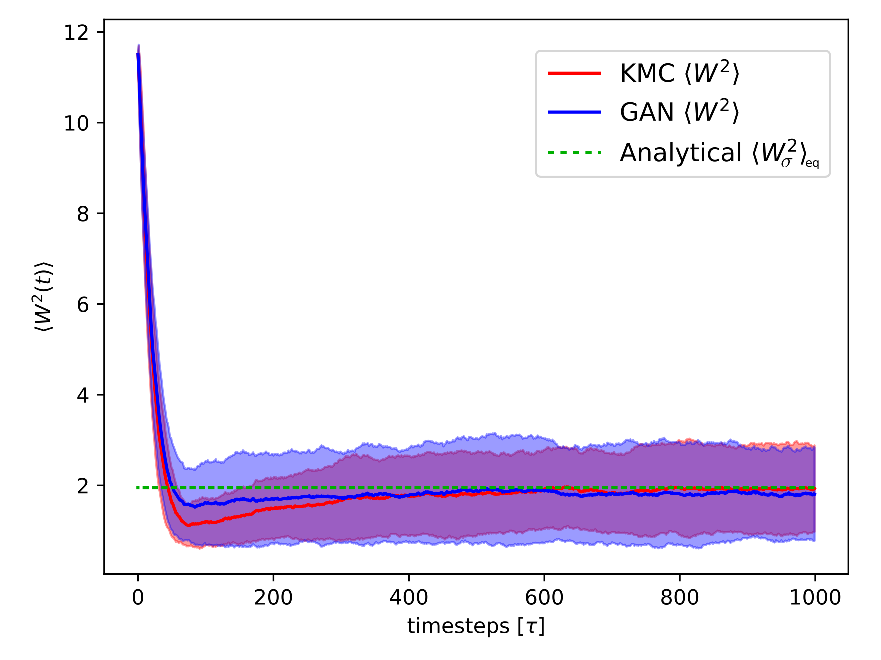}
    \caption{Comparison between GAN and KMC predictions (blue and red solid lines respectively) during relaxation dynamics from an initially sine-like stripe configuration (amplitude $5$ lattice units). The dashed green line reports the analytical equilibrium roughness value. Domain length $L=72$ and perturbation wavevector $k=4\pi/L$.}
    \label{fig::relaxation72}
\end{figure}

As a second generalization test, we also repeat the analysis of the relaxation behavior of an initially undulated configuration on a domain with $L=72$, analyzing the GAN performance in a condition combining the effects of domain size and configuration generalization. Moreover, we choose $k=4\pi/L$, so that $2$ periods are contained in the computational domain and the wavevector of the initial condition is not present in the dataset by construction. $\langle W^2(t)\rangle$ curves are reported in Fig.~\ref{fig::relaxation72}. Profile amplitude is again $A=5$ lattice units. Due to the lower curvature of the initial condition with respect to the $L=64$ case, a less pronounced overshooting of the roughness equilibrium value is present. As expected, the GAN quantitative accuracy is reduced, with the largest deviations coming from the initial relaxation procedure (the relative difference in the minimum average roughness is $\approx 35 \%$). On the other hand, the saturation value is quantitatively reproduced, with $\langle W^2\rangle_\text{GAN} = 1.83 \pm 0.04$ vs the KMC value of $1.70 \pm 0.04$ (relative error $3.4 \%$; corresponding analytical value is $1.95$).

\subsection*{Domain pinching and time extrapolation limits}

\begin{figure}
    \centering
    \includegraphics[width=\columnwidth]{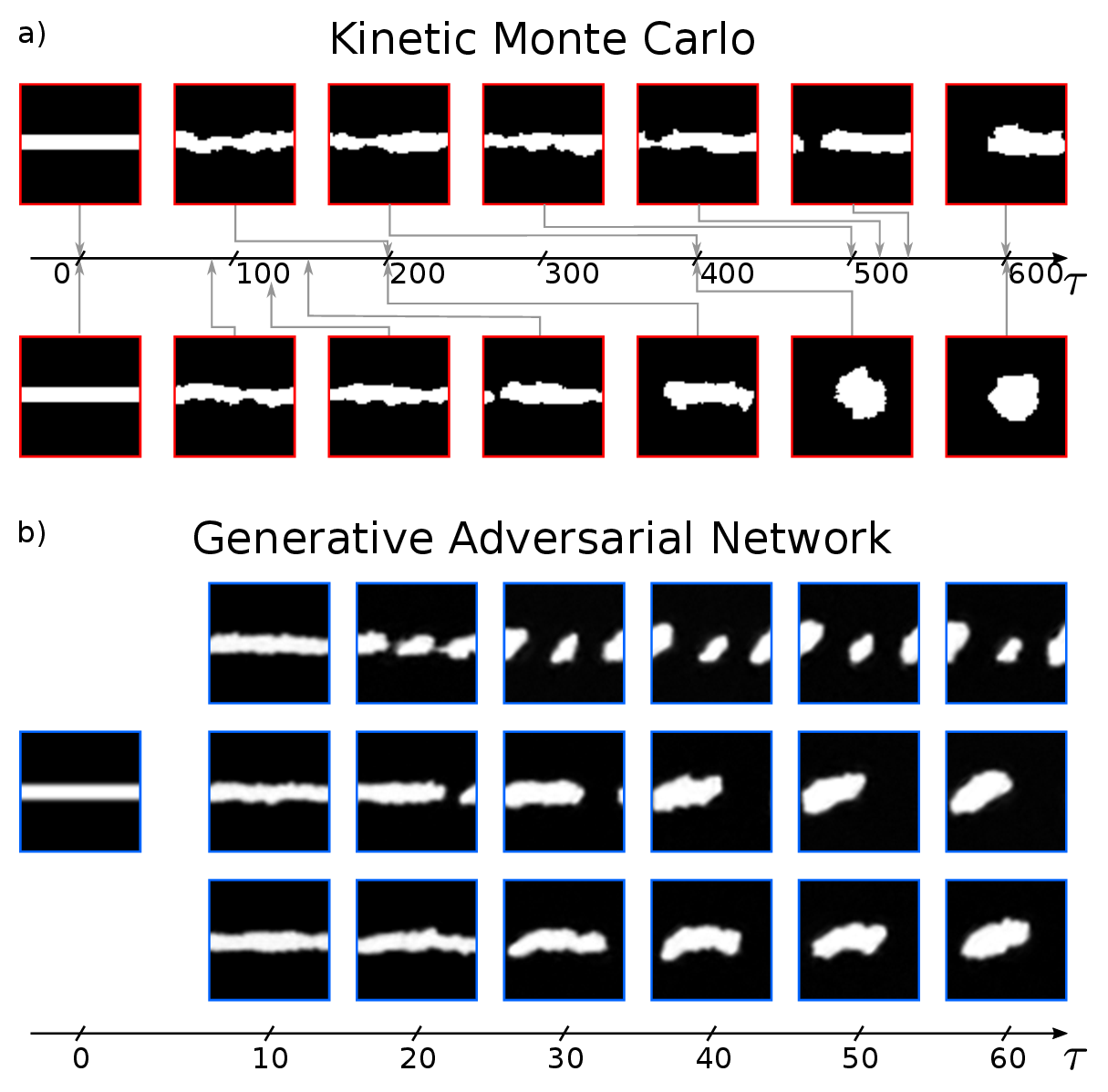}
    \caption{Example of evolutions for thin stripes undergoing pinching events due to thermal fluctuations. (a) KMC ground truth. (b) GAN predictions. The NN model qualitatively captures the behavior, but both pinching times and subsequent evolution are not compatible with the correct dynamics.}
    \label{fig::pinching}
\end{figure}

Finally, we check GAN performances in simulations where extrapolation behavior is expected, i.e., evolutions involving domain splitting. Indeed, the full dataset was built using a combination of KMC parameters and stripe sizes such that fluctuations in roughness never actually led to breaking into subdomains. By considering thinner stripes, as expected, KMC predicts pinching (see Fig.~\ref{fig::pinching}(a)). A similar behavior is obtained from GAN too, as shown in Fig.~\ref{fig::pinching}(b). Accordance, however, is only qualitative, as the typical time required to pinch the domain is much shorter with respect to the ``true'' dynamics from KMC predictions. Based on a small sample of comparable KMC simulations, GAN predictions are off by a factor of at least $\approx 10\times$, with a far less diverse distribution of pinching times if compared to the KMC ground truth. Additionally, as evidenced in Fig.~\ref{fig::pinching}(b), once the event takes place, GAN trajectories show artifacts such as the systematic diagonal elongation of islands, which are not present in late stages of Fig.~\ref{fig::pinching}(a), as expected from symmetry considerations. This is not surprising, since the NN model is operating under strong extrapolation conditions. Despite never witnessing such an event, however, it still manages to qualitatively capture a reasonable trend. To obtain quantitative predictions, however, an enlarged, specialized dataset would be required.

\section{Conclusions and perspectives} \label{sec::conclusions}

In the present work, we have shown how Generative Adversarial Networks may be effectively used to learn the dynamics of a many-particle system under the influence of thermal fluctuations. The comparison with both available analytical expressions and results from the Kinetic Monte Carlo simulations used for training the Machine Learning algorithm shows that GANs may successfully return both qualitative and quantitative results. Additionally, we tested generalization and extrapolation limits. The Generator was able to quantitatively predict scaling behavior and some details of the relaxation from strongly perturbed initial conditions outside the dataset. As expected, however, predictive capabilities are more limited and only qualitative when facing complex, unobserved phenomena that involve topological changes such as pinching. Based on this findings, it is our opinion that generative approaches may represent a new materials modeling paradigm to accelerate condensed matter simulations.

The workflow may be modified by substituting GANs with other generative models proposed in the Machine Learning community, such as Variational Autoencoders (VAE)~\cite{kingma2013auto} and diffusion models~\cite{ho2020denoising}. Considering the success the latter are recently delivering, an even higher degree of agreement with respect to the ground truth may be achievable. On the other hand, GANs are generally faster in sample generation~\cite{dhariwal2021diffusion}. Hence, the best choice for the generative strategy may depend on the physical model and on the class of phenomena being considered.

Irrespective of the generative model used, some issues still remain to be addressed and will be the topic of future studies. For example, conditioning the generated dynamics on $T$ is key and will enable the recovery of temperature-dependent behavior. From a heuristic point of view, it would make sense to correlate $T$ with the variance of $z$, so that, when $T\to0$~K, the latent space collapses to a single vector and the dynamics is effectively "frozen". Conversely, as temperature increases, larger regions of the latent space become available, allowing for broader fluctuations. A second question to be addressed is how to integrate even more physics-inspired layers into the Generator and Discriminator architectures, since these additions have already proven to highly increase the quality of NN predictions, e.g., in deterministic models of microstructure evolution~\cite{lanzoni2024extreme}.

In terms of future perspectives, the application of the presented approach to more complex dynamics, more computationally demanding for conventional simulations, is an exciting opportunity. This will allow one to collect faster and better statistics and reduce the computational bottlenecks of KMC, constituting a new paradigm for computational and condensed matter physics. Moreover, given the generality of the approach, application to other simulation methods involving thermal fluctuations, such as molecular dynamics, and in principle even experiments, is envisioned. The computational speed gain opens the perspective to close the gap with respect to experimental time-scales, similarly to continuum models such as phase field, but at the same time retaining important atomistic details and the presence of thermal fluctuations.

\section*{Data availability statement}
The data supporting the findings of this study are available upon reasonable request from the authors. The code used to train the GAN model is freely available on GitHub at~\url{https://github.com/dlanzo/DGAKMC}. The dataset used to train the model and the generated trajectories are available at \url{https://doi.org/10.24435/materialscloud:8j-b8}.

\section*{Acknowledgments}
F.M., R.B., and D.L. acknowledge financial support from ICSC—Centro Nazionale di Ricerca in High-Performance Computing, Big Data and Quantum Computing, funded by European Union—NextGenerationEU.

\clearpage

\onecolumngrid
\appendix

\section{Equivalence between 2D convolution and convolution of $h(x)$} \label{sec::convolution_appendix}
We here determine the conditions under which the 2D convolution with a gaussian kernel on the binary-image representation of the state of the system translate to a corresponding 1D convolution for the sharp step profile $h(x)$. We focus on a single edge configuration; generalization to stripes is trivial. Given a specific edge step configuration, the field $\varphi$ is simply defined as (here we neglect the possible overhangs of the profile $h(x)$):
\begin{equation}
    \varphi(x,y) = \Theta[y < h(x)],
\end{equation}
where $\Theta$ is the characteristic function, $\Theta[x] = 1$ if $x<0$ and $\Theta[x] = 0$ if $x\le 0$.

The convolution with a gaussian kernel with standard deviation $\sigma$ (assuming an infinite domain) affects the values of $\varphi$ at point $(x_0, y_0)$ as:
\begin{equation*}
\begin{split}
\varphi_\sigma(x_0, y_0) = & 
\int_{-\infty}^{\infty} \frac{\exp \left( \frac{-(x-x_0)^2}{2\sigma^2} \right)}{\sqrt{2\pi} \sigma} \int_{-\infty}^{\infty} \frac{\exp \left( \frac{-(y-y_0)^2}{2\sigma^2} \right)}{\sqrt{2\pi} \sigma} \Theta[y < h(x)] dy dx = \\
& \frac{1}{2\pi \sigma^2} \int_{-\infty}^{\infty} \exp \left( \frac{-(x-x_0)^2}{2\sigma^2} \right) \int_{-\infty}^{h(x_0)} \exp \left( \frac{-(y-y_0)^2}{2\sigma^2} \right) dx dy = \\
& \frac{1}{\sqrt{2\pi}\sigma} \int_{-\infty}^{\infty} \exp \left( \frac{-(x-x_0)^2}{2\sigma^2} \right) \frac{1}{2} \left( 1+\text{erf}\left( \frac{h(x)-y_0}{\sqrt{2}\sigma} \right) \right) dx,
\end{split}
\end{equation*}
where $\text{erf}$ is the error function. The convoluted sharp profile $h_\sigma(x)$ is implicitly defined by the $1/2$ level set of $\varphi_\sigma$:
\begin{equation*}
    \frac{1}{\sqrt{2\pi}\sigma} \int_{-\infty}^{\infty} \exp \left( \frac{-(x-x_0)^2}{2\sigma^2} \right) \frac{1}{2} \left( 1+\text{erf}\left( \frac{h(x)-h_\sigma(x_0)}{\sqrt{2}\sigma} \right) \right) dx = \frac{1}{2},
\end{equation*}
or
\begin{equation*}
    \int_{-\infty}^{\infty} \exp \left( \frac{-(x-x_0)^2}{2\sigma^2} \right) \text{erf}\left( \frac{h(x)-h_\sigma(x_0)}{\sqrt{2}\sigma} \right) dx = 0.
\end{equation*}
Due to the gaussian factor, only points with $|x-x_0| < \sigma$ contribute significantly to the integral. If
\begin{equation} \label{eq::condition5}
    \sqrt{2}\sigma \gg |h(x) - h_\sigma(x_0)|,
\end{equation}
in the neighbourhood of $x_0$, to first order
\begin{equation*}
    \int_{-\infty}^{\infty} \exp \left( \frac{-(x-x_0)^2}{2\sigma^2} \right) \frac{h(x)-h_\sigma(x_0)}{\sqrt{2}\sigma} dx \approx 0,
\end{equation*}
yielding
\begin{equation} \label{eq::condition6}
    h_\sigma(x_0) \approx \int_{-\infty}^{\infty} \frac{\exp \left( \frac{-(x-x_0)^2}{2\sigma^2} \right)}{\sqrt{2\pi} \sigma} h(x) dx,
\end{equation}
which is the same as in Ref.~\cite{gagliardi2022controlling} and allows the comparison of quantities extracted from $\varphi_\sigma$ with expressions derived from the 1D smoothed function $h_\sigma$.

The conditions under which the approximation is justified may be better specified in terms of other physical parameters. Inserting Eq.~\ref{eq::condition5} in Eq.~\ref{eq::condition6}, we have the condition:
\begin{equation} \label{eq::condition_squared}
    \left| \int_{-\infty}^{\infty} \frac{\exp \left( \frac{-(x-x_0)^2}{2\sigma^2} \right)}{\sqrt{2\pi} \sigma} [h(x_0) - h(x)] dx \right|^2 \ll 2\sigma^2.
\end{equation}
Using the Cauchy-Schwarz inequality,
\begin{equation*}
    \left| \int_{-\infty}^{\infty} \frac{\exp \left( \frac{-(x-x_0)^2}{2\sigma^2} \right)}{\sqrt{2\pi} \sigma} [h(x_0) - h(x)] dx \right|^2 \le  \int_{-\infty}^{\infty} \frac{\exp \left( \frac{-(x-x_0)^2}{2\sigma^2} \right)}{\sqrt{2\pi} \sigma} \left| h(x_0) - h(x) \right|^2  dx.
\end{equation*}
Using the equilibrium ensemble average for $ \left| h(x_0) - h(x) \right|^2 = k_BT|x-x_0|/\tilde{\gamma}$~\cite{misbah2010crystal}, we obtain:
\begin{equation*}
    \int_{-\infty}^{\infty} \frac{\exp \left( \frac{-(x-x_0)^2}{2\sigma^2} \right)}{\sqrt{2\pi} \sigma} \frac{k_BT}{\tilde{\gamma}} \left| h(x_0) - h(x) \right|  dx = \sqrt{\frac{2}{\pi}} \frac{k_B T}{\tilde{\gamma}} \sigma,
\end{equation*}
leading to the sufficient condition:
\begin{equation}
\sqrt{\frac{2}{\pi}} \frac{k_B T}{\tilde{\gamma}} \ll 2 \sigma,
\end{equation}
which is verified for the parameters used in the work, since $\sqrt{2/\pi} k_B T/\tilde{\gamma} \approx 0.29$ and $2\sigma \approx 2.22$.

As a final condition, since the microscopic non-averaged profiles $h(x)$ exhibit height jumps of the order of the lattice spacing $a$, we must require the condition $\sigma \gtrapprox a$, which again is verified in our case. To be more precise, consider the special case of a simple profile of kink height $h(x)=h_0$ for $x \le x_0$ and $h(x) = h_0+a$ for $x > x_0$. Then Eq.~\ref{eq::condition_squared} leads to:
\begin{equation}
    \left( \frac{a}{2} \right)^2 \ll 2 \sigma^2,
\end{equation}
leading to the relationship $a \ll 2\sqrt{2} \sigma$.

\section{Derivation of roughness variance} \label{sec::roughness_derivation}

To obtain an analytical expression for the variance of the equilibrium roughness, we define for every time $\xi(x)$ as the deviation of the step edge with respect to its average position:
\begin{equation} \label{eq::xi_definition}
    \xi(x) = h(x) - \bar{h},
\end{equation}
with $\bar{h} = \int_0^Lh(x)dx/L$ and $x\in[0,L]$ as in the main text. From the definition of roughness Eq.~\ref{eq::roughness_definition}, it follows that:
\begin{equation*}
    W^2 = \frac{1}{L} \int_0^L\xi^2(x) dx.
\end{equation*}

We define the roughness variance $\mu_2$ as:
\begin{equation} \label{eq::roughvariance_definition}
    \mu_2 = \langle W^4 \rangle - \langle W^2 \rangle^2.
\end{equation}
Using the definition Eq.~\ref{eq::xi_definition} and Eq.~\ref{eq::roughvariance_definition}:
\begin{equation*}
    \mu_2 = \frac{1}{L^2} \iint \langle \xi^2(x)\xi^2(x') \rangle - \langle \xi^2(x) \rangle \langle \xi^2(x') \rangle dx dx'.
\end{equation*}
Writing $\xi$s in terms of their Fourier transforms $\xi(q) = \int_0^L h(x) \exp \left(-iqx\right) dx$:
\begin{equation*}
\begin{split}
    \mu_2 = & \frac{1}{L^2} \iint \prod_{j=1}^4 \int e^{i(q_1+q_2)x} e^{i(q_3+q4)x'} \left[ \langle \xi(q_1)\xi(q_2)\xi(q_3)\xi(q_4) \rangle - \langle \xi(q_1)\xi(q_2) \rangle \langle \xi(q_3)\xi(q_4) \rangle \right] \frac{dq_j}{2\pi} dx dx' = \\
    & \frac{1}{L^2} \iint \prod_{j=1}^4 \int e^{i(q_1+q_2)x} e^{i(q_3+q4)x'} \left[ \langle \xi(q_1)\xi(q_3)\rangle \langle \xi(q_2)\xi(q_4) \rangle - \langle \xi(q_1)\xi(q_4) \rangle \langle \xi(q_2)\xi(q_3) \rangle \right] \frac{dq_j}{2\pi} dx dx',
\end{split}
\end{equation*}
where the second equality follows from Wick's theorem. For a periodic system with periodicity in $[0,L]$, we have $q_j=2\pi n_j/L$ and $\int dq_i/2\pi \to 1/L \sum_{n_j}$, where $n_j \in \mathbb{Z}$ indicize normal modes. Additionally, we substitute the Fourier transform with discrete Fourier components $\xi(q_i) \to \xi_{q_i}$. Integrating with respect to $x$ and $x'$:
\begin{equation*}
    \mu_2 = \frac{1}{L^4} \sum_{n_1} \sum_{n_3} \left[ \langle \xi_{q_1}\xi_{q_3} \rangle  \langle \xi_{-q_1}\xi_{-q_3} \rangle - \langle \xi_{q_1}\xi_{-q_3} \rangle \langle \xi_{-q_1}\xi_{q_3} \rangle \right].
\end{equation*}
At equilibrium, Fourier modes are orthogonal, i.e. $\langle \xi_{q_i}\xi_{q_j} \rangle_\text{eq} = \langle |\xi_{q_i}|\rangle_\text{eq} \delta_{n_i+n_j}$, hence:
\begin{equation*}
    \mu_2 = \frac{1}{L^4} \sum_{n_1} \sum_{n_3} \left[ \delta_{n_1+n_3} \langle \xi_{q_1} \xi_{-q_1} \rangle^2_\text{eq} + \delta_{n_1-n_3} \langle \xi_{q_1}\xi_{-q_1} \rangle^2_\text{eq} \right],
\end{equation*}
where we used the relation $\delta_n^2=\delta_n$. Summing over $n_3$:
\begin{equation*}
    \mu_2 = \frac{1}{L^4} \sum_{n} 2 \langle \xi_{q}\xi_{-q} \rangle^2_\text{eq} = \frac{2}{L^4} \sum_{n} \langle |\xi_{q}| \rangle^2_\text{eq},
\end{equation*}
where we dropped the $1$ subscript to lighten the notation. Using equipartition for $q \ne 0$
\begin{equation}
    \langle |\xi_q| \rangle^2_\text{eq} = \frac{k_BTL}{\tilde{\gamma}q^2} = \frac{k_BT L^3}{\tilde{\gamma}(2\pi n)^2},
\end{equation}
hence
\begin{equation}
    \mu_2 = \frac{2}{L^4} \sum_{n\ne 0} \left( \frac{k_BT L^3}{\tilde{\gamma} (2\pi n^2)} \right)^2 = 2 \left( \frac{k_BT}{\tilde{\gamma}} \right)^2 \frac{L^2}{(2\pi)^4} \sum_{n \ne 0} \frac{1}{n^4} = 2 \left( \frac{k_BT}{\tilde{\gamma}} \right)^2 \frac{L^2}{(2\pi)^4} \frac{\pi^4}{90} = \frac{L^2}{360} \left( \frac{k_BT}{\tilde{\gamma}} \right)^2.
\end{equation}

Finally, taking the ratio between the roughness standard deviation $\sqrt{\mu_2} = \text{std}(W^2)_\text{eq}$ and its mean equilibrium value $\langle W^2 \rangle_\text{eq}$ Eq.~\ref{eq::eq_roughness_noconv} we obtain
\begin{equation}
    \frac{\sqrt{\mu_2}}{\langle W^2 \rangle_\text{eq}} = \sqrt{\frac{2}{5}}.
\end{equation}

\hspace{10pt}
\twocolumngrid
\bibliography{biblio}

\end{document}